 \documentclass[
twocolumn, amsmath, amssymb, floatfix,
nofootinbib,wrapfloat byrevtex]{revtex4}

 \usepackage{amsmath,mathrsfs,graphicx,epstopdf,placeins,float}
 \usepackage{booktabs}
 \usepackage{multirow}
 \usepackage{pstricks}

 \newcommand{\beq}[1]{\begin{equation}\label{#1}}
 \newcommand{\eeq}{\end{equation}}
 \newcommand{\bea}[1]{\begin{eqnarray}\label{#1}}
 \newcommand{\eea}{\end{eqnarray}}

\begin{document} 

 \title{Counterterm method and thermodynamics of Hairy Black Holes in a Vector-Tensor theory with Abelian gauge symmetry breaking }
 
 \author{Ai-chen Li $^{a,b}$}
\email{lac@emails.bjut.edu.cn}
 \affiliation{\it ${}^a$ Departamento de matematica da Universidade de Aveiro and CIDMA,\\ Campus de Santiago, 3810-183 Aveiro, Portugal} 
 \affiliation{\it ${}^b$ Theoretical Physics Division, College of Applied Sciences, Beijing University of Technology }

 \begin{abstract}
 For a type of non-minimally coupled vector-tensor theories with Abelian gauge symmetry breaking in four-dimensional spacetime and correspondingly asymptotic non-AdS black hole solutions including a cosmological constant, we construct the appropriate boundary terms and derive the associated junction condition. In order to remove the divergences in the stress tensor which is localized on the spacetime boundary, we also involve the suitable surface counterterms into the total action. Using the counterterm method, we caculate the black hole mass. An implicit relation between the black hole carge $Q$ and other parameters is implied by combining the expression of the black hole mass with the first law of black hole thermodynamics. With this implicit relation, we can prove the inequality $Q\leq M$ which is a general bound for most of charged black holes. Besides, the phase structure of black holes is also investigated in the grand canonical ensemble.
 \end{abstract}
\maketitle

\section{Introduction}

In the recent decades, the studies about hairy black holes have become a hot topic. One  motivation is the Anti de-Sitter space/conformal field theory correspondence \cite{Maldacena:1997re, Gubser:1998bc, Witten:1998qj}, i.e. AdS/CFT. This correspondence suggests that the physics of strongly coupled gauge field living on  boundary of AdS spacetime could be reproduced by classical or semi-classical gravitational theory in the bulk. In application of AdS/CFT, the hairy black holes are used to understand quantum chromodynamics (QCD) at finite temperature \cite{Gursoy:2008bu,Gursoy:2008za,Gubser:2008yx},  superconductor's behavior \cite{Hartnoll:2008vx, Hartnoll:2008kx}, and other interesting phase transitions \cite{Hartnoll:2007ai, Hartnoll:2007ip, Astefanesei:2019ehu, Hennigar:2016xwd, Xu:2019yub,Mahapatra:2020wym}. Besides, another motivation comes from the re-examination to the black hole no-hair theorems. According to them, in four-dimensional spacetime with gravity, black holes are characterized only by three physical parameters, namely the mass, electric charge, and angular momentum. And the existence of black holes characterized by other parameters are excluded. However, some interesting counterexamples have been constructed in recent years see $e.g.$ \cite{Cai:1997ij, Martinez:2002ru, Zloshchastiev:2004ny, Winstanley:2005fu, Brihaye:2005an, Brihaye:2006kn, Herdeiro:2014goa, Herdeiro:2014ima, Benone:2014ssa, Brihaye:2014nba,Herdeiro:2016tmi,Antoniou:2017acq,Antoniou:2017hxj}. Moreover, recently the phenomenon of black hole spontaneous scalarization has been the focus of considerable attention - see $e.g.$ \cite{Silva:2017uqg, Doneva:2017bvd, Doneva:2018rou, Minamitsuji:2018xde, Silva:2018qhn, Macedo:2019sem, Cunha:2019dwb, Herdeiro:2019yjy, Herdeiro:2018wub}. Specifically, these models consider a non-minimal coupling between the scalar field $\phi$ and some source therm $I$, which could produce a repulsive gravitational effect. In this way, the black hole solution with no hair, from Einstein's gravity, is unstable against scalar perturbations due to the source, and the scalar (or more general vector or tensor) hair will grow dynamically during this process.

Most hairy black hole solutions have been considered in scalar-tensor theories or scalar-vector-tensor theories. Solutions from vector-tensor theories have been less investigated. Actually, many interesting cosmological phenomenology have been found in vector-tensor theories, especially for those with Abelian symmetry breaking \cite{Golovnev:2008cf, BeltranJimenez:2010uh, EspositoFarese:2009aj, Tasinato:2014eka, Tasinato:2014mia, Mann:2011hg, Herdeiro:2011uu, Ovgun:2017hje, DeFelice:2016yws, Zhou:2017glv, Deng:2018wrd}. Recently, a class of interesting black hole solutions in four-dimensional spacetime were obtained by \cite{Chagoya:2016aar} from a type of vector-tensor theory in which the Abelian gauge symmetry is broken by coupling the vector with gravity in a non-minimal way~\cite{Gripaios:2004ms, Heisenberg:2014rta}.

In this paper, our purpose is to find the appropriate boundary terms and construct the effective holographic surface counterterms for a type of non-minimally coupled vector-tensor theory and correspondingly asymptotic non-AdS black hole solutions including a cosmological constant \cite{Chagoya:2016aar}. After obtaining the black hole mass via a surface counterterms method, we also intend to investigate the relevant thermodynamics. Our motivation comes from the following aspects, both in physics and in  techniques. As we know, for the Reissner-Nordstr$\ddot{o}$m black hole, the value of the charge $Q$ should be less than the value of the black hole mass $M$; otherwise no event horizon exists. More generally speaking, the extremality bound $Q\leq M$ holds for most of charged black hole solutions, of course, there also exists some interesting counterexamples\cite{Herdeiro:2015gia, Delgado:2016zxv}). However, for black hole solutions including a cosmological constant in \cite{Chagoya:2016aar}, the value of $Q$ is arbitrary and has no effect on the existence of horizon. It seems that there is no direct evidence to show $Q\leq M$ or $Q>M$ or both are allowed. Thus, after the consideration of thermodynamics for these black hole solutions, we expect to find a constraint relation between $Q$ and other physical parameters. In this way, we can compare $Q$ and $M$ at other fixed physical parameters. Furthermore, the thermodynamics for black hole solutions derived from some typical vector-tensor theories with broken Abelian gauge symmetry has been investigated by works \cite{Fan:2017bka, Liu:2014tra}, via the Wald formalism. These works derive the general first laws of black hole thermodynamics in either minimally or non-minimally coupled case. But the physics behind these formulas have not been thoroughly investigated. In this work, we expect to gain some intuition about the black holes in vector-tensor theory without Abelian gauge symmetry~\cite{Chagoya:2016aar} by analysing the phase structures in the grand canonical ensemble.

From the viewpoint of techniques, due to the presence of a non-minimal coupling between the vector field and gravity in the action given in \cite{Chagoya:2016aar}, the Gibbons-Hawking term~\cite{Gibbons:1976ue} is not the appropriate boundary term here. The effective boundary term needs to be obtained. Besides, for obtaining a finite  quasilocal stress tensor on spacetime boundary, some suitable surface counterterms should be added to the total action. This process is the so called holographic renormalization~\cite{Henningson:1998gx, Hyun:1998vg, Chalmers:1999gc, Nojiri:1998dh, Balasubramanian:1999re, Emparan:1999pm, deHaro:2000vlm}. However, in \cite{Chagoya:2016aar}, there exists non-minimal coupling between vector field and gravity, while the black hole solution is not an exactly asymptotic AdS geometry. Thus, the surface counterterms will be more complicate than \cite{Balasubramanian:1999re}. At present, there are few research works about resolving these technical problems. Thus, to enrich the studies about the vector-tensor theories with Abelian symmetry breaking, we shall consider these technical problems for \cite{Chagoya:2016aar} in this work.
    
Our work is structured as follows. In Sec.\ref{fieldeq}, we briefly review a type of vector-tensor theory with broken Abelian gauge symmetry in four-dimensional spacetime and correspondingly asymptotic non-AdS black hole solutions including a cosmological constant. In Sec.\ref{calJunCou}, we construct the appropriate boundary terms and derive the corresponding generalised Israel junction conditions on the spacetime boundary. Besides, we also give a well-defined  quasilocal stress tensor and calculate the black hole mass by adding the suitable surface counterterms to the total action. After obtaining the black hole mass, the relevant thermodynamics, including the first law of black hole thermodynamics and the phase structure analysis in the grand canonical ensemble, have been considered in Sec.\ref{BHThermodynamic}. Finally, we will summarize our results and give a discussion in Sec.\ref{ConDiss}. 

\section{Black Holes solution including a cosmological constant in vector-tensor theory with Abelian symmetry breaking \label{AdSProBHSol}}
\label{fieldeq}

An analytical black hole solution in 4-dimensional spacetime has been found in a type of vector-tensor theory \cite{Chagoya:2016aar}. The action is set as
\begin{align}
\label{EinProBHaction}
\hspace{-2mm}S=\frac{1}{2\kappa^{2}}\int\sqrt{-g}d^{4}x\big\{ R-2\Lambda-\frac{1}{4}F^{2}+\beta G_{\mu\nu}A^{\mu}A^{\nu}\big\}
\end{align}
in which the $G_{\mu\nu}$ is the standard Einstein tensor, while the $\beta$ is the physical constant which measures the strength of nonminimal coupling between the vector field and Einstein tensor. It is easy to see that the $U(1)$ symmetry is broken in presence of this nonminimal coupling term. The Einstein field equation is given by
\begin{align}
\nonumber
&\quad \quad R_{\mu\nu}-\frac{1}{2}g_{\mu\nu}R=-\Lambda g_{\mu\nu}+\beta Z_{\mu\nu}\\
\label{EinFieEqua}
&\quad \quad\quad \quad\quad\quad\quad\quad+\frac{1}{2}(F_{\mu\sigma}F_{\nu}^{\sigma}-\frac{1}{4}g_{\mu\nu}F^{2})\\
\nonumber
\\
\nonumber
&Z_{\mu\nu}=\frac{1}{2}A^{2}R_{\mu\nu}+\frac{1}{2}RA_{\mu}A_{\nu}-2A^{\alpha}R_{\alpha(\mu}A_{\nu)}-\frac{1}{2}\nabla_{\mu}\nabla_{\nu}A^{2}\\
\nonumber
&\quad \quad +\nabla_{\alpha}\nabla_{(\mu}(A_{\nu) }A^{\alpha})-\frac{1}{2}\nabla^{\alpha}\nabla_{\alpha}(A_{\mu}A_{\nu})\\
\nonumber
&\quad\quad+\frac{1}{2}g_{\mu\nu}\big(G_{\alpha\beta}A^{\alpha}A^{\beta}+\nabla^{\alpha}\nabla_{\alpha}A^{2}-\nabla_{\alpha}\nabla_{\beta}(A^{\alpha}A^{\beta})\big)
\end{align}
the equation of motion for the vector field, namely the extended Maxwell equations, reads
\begin{align}
\label{procaequa}
\nabla_{\mu}F^{\mu\nu}+2\beta A_{\mu}G^{\mu\nu}=0
\end{align}
The metric ansatz is set up as
\begin{align}
\label{MetAnsatz}
&ds^2=-h(r)dt^2 +\frac{dr^2}{f(r)}+r^2(d\theta^2+\sin^2\theta d\phi^2) 
\end{align}
Meanwhile, $A_\mu$ has the following ansatz
\begin{align}
\label{ElecPotenAnsatz}
A_{\mu}dx^{\mu}=a(r)dt+\chi(r)dr
\end{align}
After substituting ansatz $\eqref{MetAnsatz}$ and $\eqref{ElecPotenAnsatz}$ into Einstein field equations $\eqref{EinFieEqua}$ and extend Maxwell equations $\eqref{procaequa}$, the following four indepentent differential equations are given \cite{Chagoya:2016aar}
\begin{align}
\label{Vecr}
&\hspace{47.4mm}\frac{f^{2}}{r}-\frac{f}{r}+\frac{h^{\prime}f^{2}}{h}=0\\
\label{Vect}
&\hspace{-1.5mm}4\beta a(1-f-rf^{\prime})+a^{\prime}(r-5rf-r^{2}f^{\prime})-2r^{2}a^{\prime\prime}f=0\\
\label{Einrr}
&\hspace{-4.2mm}4\beta\big(a^{2}(f-1)+f(h\chi^{2}+2raa^{\prime})\big)+r^{2}(4\Lambda h+{a^{\prime}}^2f)=0\\
\nonumber
&\hspace{0.1mm}2\beta\big(a^{2}(1-f)+\chi^{2}fh(1+f)\big)+4h(f+r^{2}\Lambda-1)+\\
\label{Eintt}		
&\hspace{-4.9mm}
r^{2}f(a^{\prime2}+\frac{4hf^{\prime}}{rf})+2r\beta f^{2}\chi^{2}(\frac{3hf^{\prime}}{f}-\frac{a^{2}f^{\prime}}{\chi^{2}f^{2}}+\frac{4h\chi^{\prime}}{\chi})=0
\end{align}
When $\beta=\frac{1}{4}$, an analytical solution of spherically symmetric black hole is given as
\begin{align}
\label{metrictt}
&h(r)=1-\frac{2m}{r}+\frac{4r^{2}\Lambda_{eff}}{3}+\frac{4}{5}r^{4}\Lambda_{eff}^{2}\\
\label{metricrr}
&f(r)=(1+2r^{2}\Lambda_{eff})^{-2}h(r)\\
\label{elect}
&a(r)=\frac{Q}{r}+Q_{2}(1+\frac{2}{3}r^{2}\Lambda_{eff})\\
\nonumber
&\chi(r)=\frac{1+2r^{2}\Lambda_{eff}}{\sqrt{h}}\big(\frac{(3Q+rQ_{2}(3+2r^{2}\Lambda_{eff}))^{2}}{9r^{2}h}\\
\label{elecr}
&\quad\quad-Q_{2}^{2}(1+2r^{2}\Lambda_{eff})+16r^{2}\Lambda_{eff}\big)^{\frac{1}{2}}
\end{align}
where 
\begin{align}
\label{EffLambda}
\Lambda_{eff}=\frac{2\Lambda}{Q_{2}^{2}-8}
\end{align}
From $\eqref{metrictt}$, we see that the effective cosmological constant in the black hole solutions is $\eqref{EffLambda}$. Thus, for obtaining an effective negative cosmological constant, we will only consider the case $Q_2 ^2 >8$ in Sec.\ref{calJunCou} and Sec.\ref{BHThermodynamic}. Actually, one could also consider the case $\Lambda>0~\&~Q^2_2<8$ as an alternative way. And the same results will be observed. Besides, note that the asymptotic behavior of black hole solutions at large $r$ is
\begin{align}
\label{AsyLarger}
ds^{2}\sim-\frac{4\Lambda_{eff}^{2}}{5}r^{4}dt^{2}+\frac{1}{5}dr^{2}+r^{2}(d\theta^2+\sin^2\theta d\phi^2)
\end{align}
It's easy to see that this black hole solutions including a cosmological constant in \cite{Chagoya:2016aar} are not asymptotatic AdS geometry.

\begin{figure}[ht]
	\begin{center}
		\includegraphics[scale=0.33]{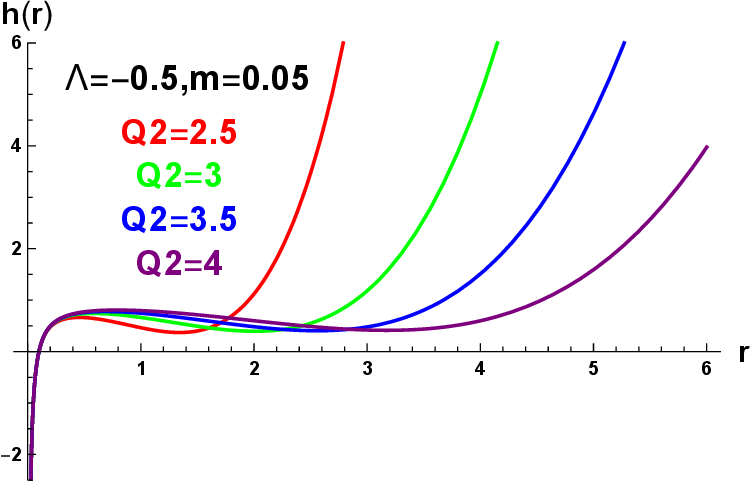}
		\includegraphics[scale=0.33]{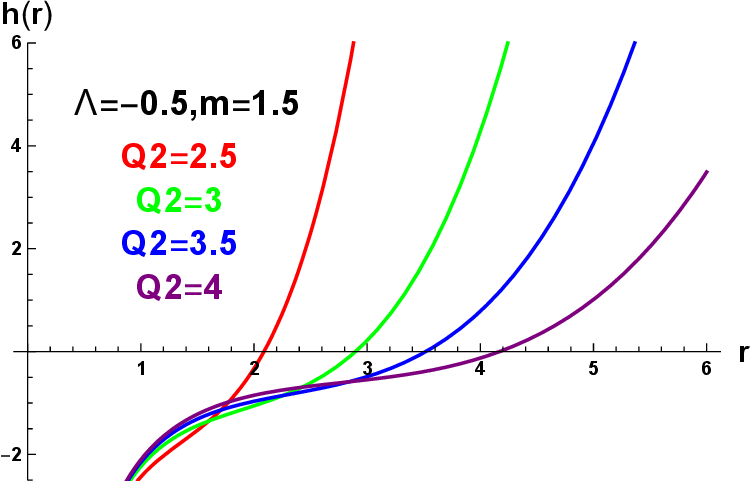}
		\caption{(color online).Plot the horizon function $h(r)$ \eqref{metrictt} at fixed $\Lambda, m$ with different $Q_2$. }
		\label{numhorizon}
	\end{center}
\end{figure}

We plot $h(r)$ in fig.\ref{numhorizon} at some fixed physical parameters $\Lambda,m$ with varying $Q_2$. It's easy to see that there only exists one horizon $r_h$ for this black hole solution whatever the value of $\Lambda, m, Q_2$.
Besides, from $\eqref{metrictt}$-$\eqref{elecr}$, it is not difficult to observe that the value of $Q$ is arbitrary with respect to other parameters and has no effect on the existence of a horizon $r_h$.

\section{Junction condition, counterterm method and holographic energy
\label{calJunCou}}

In this part, we need to construct some physical quantities on the spacetime boundary $r= \infty$. Before that, some notation should be introduced. We use $x^\mu=(t,r,\theta,\phi)$ and $ds^2=g_{\mu\nu}dx^\mu dx^\nu$ to denote the coordinates and metric in the  bulk spacetime. In the spacetime boundary, namely $r=\infty$, the coordinates are denoted as $x^{a}=(t,\theta,\phi)$. Thus, the 4-velocity of the boundary hypersurface is easily obtained as $u^{\mu}=(1,0,0,0)$, and the unit normal pointing into the boundary hypersurface is $n_{\nu}=(0,\frac{1}{\sqrt{f}},0,0)$. Then the induced metric of the boundary spacetime could be obtained as
\begin{align}
\label{indumetric}
&ds^2=\gamma_{ab}dx^a dx^b=-h(r)dt^{2}+r^{2}(d\theta^{2}+\sin^{2}\theta d\phi^{2})\\
\nonumber
&\gamma_{ab}=e^\mu _a e^\nu _b \gamma_{\mu\nu} 
\end{align}
where the vielbein $e^\mu _a$ is defined as $e^\mu _a=\frac{\partial x^\mu}{\partial x^a}$, which is tangent to the  boundary and satisfies $n_\mu e^\mu _a =0$. Besides, the projection tensor $\gamma_{\mu\nu}$ is defined as $\gamma_{\mu \nu}=g_{\mu \nu}-n_\mu n_\nu$, whose tangential components $\gamma_{ab}$ correspond to the induced metric on spacetime the boundary. For caculating the quasilocal stress tensor on spacetime the boundary, we need to add the boundary terms to the action $\eqref{EinProBHaction}$. As it is known, when varying the Einstein-Hilbert action with respect to the metric tensor $g_{\mu \nu}$, besides a bulk term which yields the standard Einstein field equation, we also obtain the following boundary term
\begin{align}
\label{bounEinHilbert}
&\frac{1}{2\kappa^{2}}\int\sqrt{-\gamma}d^{3}x\big\{ g_{\alpha\beta}n_{\rho}\nabla^{\rho}(\delta g^{\alpha\beta})-n_{\rho}\nabla_{\lambda}(\delta g^{\rho\lambda})\big\}
\end{align}
in which the $n_\rho$ is the unit normal vector pointing into the spacetime boundary. Note that in expression of $\eqref{bounEinHilbert}$, the derivative of the metric variation in the normal direction is discontinuous across the boundary hypersurface. For cancelling the $\nabla^\rho (\delta g^{\alpha \beta})$-like terms and making the metric variation well-defined on the spacetime boundary, a Gibbons-Hawking boundary term \cite{Gibbons:1976ue} is involved
\begin{align}
S_{GH}=\frac{1}{\kappa^2}\int d^3x \sqrt{-\gamma} K
\end{align}
where $K=\gamma^{\mu\nu}K_{\mu\nu}$ is the trace of the extrinsic curvature tensor $K_{\mu \nu}=\frac{1}{2}(\nabla_\mu n_\nu +\nabla_\nu n_\mu)$. Adding the $\eqref{bounEinHilbert}$ with the variation of the Gibbons-Hawking term together, we will obtain the following Israel junction term \cite{Israel:1966rt, Chamblin:1999ya, Li:2018jxy}
\begin{align}
\label{variEHandGH}
&\hspace{-5.7mm}\frac{1}{\kappa^2}\int \sqrt{-\gamma} d^3x \big\{ D^\nu(\gamma^{\alpha}_\nu n^\beta \delta g_{\alpha \beta})+ (K_{\mu\nu}-h_{\mu \nu}K)\delta g^{\mu \nu} \big\}
\end{align}
in which $D^\nu$ is the covariant differentiation with respect to $\gamma_{\mu\nu}$. Thus, the first term in \eqref{variEHandGH} is a total derivative and could be thrown away. In the case of Einstein-Proca theory \eqref{EinProBHaction}, besides the Einstein-Hilbert action, the nonminimal coupling $\frac{1}{2\kappa^2} \int \sqrt{-g}d^4x \beta G_{\mu\nu}A^\mu A^\nu$ will also lead to a pathological boundary term when calculating the variation with respect to $g^{\mu \nu}$. Thus we also need to introduce an appropriate boundary term to make the nonminimal coupling term have a well-defined behavior on the spacetime boundary. 

\subsection{An appropriate boundary term which corresponds to the non-minimal coupling terms}

Varying the term $\frac{1}{2\kappa^2} \int \sqrt{-g}d^4x \{\beta G_{\mu\nu}A^\mu A^\nu \}$ with respect to the  metric $g^{\mu \nu}$, besides a bulk term $Z_{\mu \nu}$ in $\eqref{EinFieEqua}$, gives the following boundary term
\begin{align}
\nonumber
&\frac{\beta}{2\kappa^{2}}\int d^{3}x\sqrt{-\gamma}\big\{ \frac{1}{2}n_{\rho}\nabla^{\rho}(\delta g^{\alpha\beta})A_{\alpha}A_{\beta}-A^{\mu}\nabla_{\mu}(\delta g^{\lambda\rho})A_{\lambda}n_{\rho}\\
\nonumber
& \quad \quad \quad \quad+\frac{1}{2}n_{\nu}A^{\nu}A^{\mu}\nabla_{\mu}(\delta g^{\lambda\rho})g_{\lambda\rho}+\frac{1}{2}A^{2}\nabla_{\lambda}(\delta g^{\lambda\rho})n_{\rho}\\
\nonumber
& \quad \quad \quad \quad -\frac{1}{2}A^{2}n^{\alpha}\nabla_{\alpha}(\delta g^{\lambda\rho})g_{\lambda\rho}+\delta g^{\lambda\rho}\nabla_{\rho}(A_{\lambda}A^{\nu})n_{\nu}\\
\nonumber
& \quad \quad \quad \quad-\frac{1}{2}\delta g^{\alpha\beta}n^{\rho}\nabla_{\rho}(A_{\alpha}A_{\beta})-\frac{1}{2}g_{\lambda\rho}\delta g^{\lambda\rho}n_{\mu}\nabla_{\nu}(A^{\mu}A^{\nu})\\
\label{bounEinProca}
& \quad \quad \quad \quad-\frac{1}{2}n_{\lambda}\delta g^{\lambda\rho}\nabla_{\rho}A^{2}+\frac{1}{2}\delta g^{\lambda\rho}g_{\lambda\rho}n^{\alpha}\nabla_{\alpha}A^{2} \big\}
\end{align}
Note that there also exists the $\nabla^\mu (\delta g^{\alpha \beta})$-like terms in $\eqref{bounEinProca}$, which are discontinuous across the boundary hypersurface. For obtaining the well-defined junction condition on the spacetime boundary, we involve the following boundary term
\begin{align}
\nonumber
S_{sur}^\beta=\frac{\beta}{2\kappa^{2}}\int d^{3}x\sqrt{-\gamma}\big\{ -A^{\nu}(\nabla_{\nu}A^{\rho})n_{\rho}\\
\label{EinProcaBoun}
\quad \quad+n_{\nu}A^{\nu}(\nabla_{\rho}A^{\rho})-A^{2}g^{\alpha\beta}(\nabla_{\beta}n_{\alpha})\big\}
\end{align}
\label{TotalBou}
Thus the total boundary term is given as,
\begin{align}
S_{sur}=S_{GH}+S_{sur}^\beta
\end{align}
The variation of $S_{sur}^\beta$ with respect to $g^{\mu \nu}$ is
\begin{align}
\nonumber
&\frac{\beta}{2\kappa^{2}}\int d^{3}x\sqrt{-\gamma}\big\{-A^{\nu}n_{\rho}\delta\Gamma_{\nu\alpha}^{\rho}A^{\alpha}+n_{\nu}A^{\nu}(\delta\Gamma_{\rho\sigma}^{\rho}A^{\sigma})\\
\nonumber
&\hspace{-5mm}\quad\quad~-\frac{1}{2}A^{2}(\delta\Gamma_{\alpha\beta}^{\alpha}n^{\beta})+\frac{1}{2}A^{2}g^{\alpha\beta}(\delta\Gamma_{\beta\alpha}^{\rho}n_{\rho})+n_{\nu}A^{\nu}\delta g^{\rho\sigma}(\nabla_{\rho}A_{\sigma})\\
\nonumber
&\hspace{-5mm}\quad\quad~+\frac{1}{2}\gamma_{\mu\nu}\delta g^{\mu\nu}A^{\alpha}(\nabla_{\alpha}A^{\rho})n_{\rho}-\frac{1}{2}\gamma_{\alpha\beta}\delta g^{\alpha\beta}n_{\nu}A^{\nu}(\nabla_{\rho}A^{\rho})\\
\nonumber
&\hspace{-5mm}\quad\quad~+\frac{1}{2}\gamma_{\mu\nu}\delta g^{\mu\nu}A^{2}g^{\alpha\beta}(\nabla_{\beta}n_{\alpha})-\delta g^{\mu\nu}A_{\mu}A_{\nu}g^{\alpha\beta}(\nabla_{\beta}n_{\alpha})\\
\label{DelEinProcaBoun}
&\hspace{-5mm}\quad\quad~-A^{2}\delta g^{\alpha\beta}(\nabla_{\beta}n_{\alpha})-\delta g^{\mu\nu}A_{\mu}(\nabla_{\nu}A^{\rho})n_{\rho}\big\}
\end{align}  
in which the $\delta\Gamma_{\mu\nu}^{\rho}$ is defined as 
\begin{align}
\nonumber
\hspace{-3mm}\delta\Gamma_{\mu\nu}^{\rho}=-\frac{1}{2}\big(g_{\lambda\mu}\nabla_{\nu}(\delta g^{\lambda\rho})+g_{\lambda\nu}\nabla_{\mu}(\delta g^{\lambda\rho})-g_{\mu\alpha}g_{\nu\beta}\nabla^{\rho}(\delta g^{\alpha\beta})\big)
\end{align}
Adding the $\eqref{variEHandGH}$ with $\eqref{bounEinProca},\eqref{DelEinProcaBoun}$ together, the quasilocal stress tensor on spacetime boundary could be given by the following junction condition
\begin{align}
\nonumber
&T_{ab}=K_{ab}-K\gamma_{ab}+\beta\big(\frac{1}{2}\gamma_{ab}A^{2}K+\frac{1}{2}\gamma_{ab}n^{\rho}\nabla_{\rho}A^{2}\\
\nonumber
&\quad ~ -A^{2}K_{ab}-\gamma_{ab}n_{\alpha}A^{\alpha}\nabla_{\beta}A^{\beta}+2e_{a}^{\mu}e_{b}^{\nu}n_{\alpha}A^{\alpha}\nabla_{\mu}A_{\nu}\\
\label{ExtenJuncCon}
&\quad ~-e_{a}^{\mu}e_{b}^{\nu}n^{\rho}\nabla_{\rho}A_{\mu}A_{\nu}-e_{a}^{\mu}e_{b}^{\nu}A_{\mu}A_{\nu}K\big)
\end{align}
where $K_{ab}=e^\mu _a e^\nu _b K_{\mu \nu} $ and $\gamma_{ab}=e^\mu _a e^\nu _b \gamma_{\mu\nu}=e^\mu _a e^\nu _b g_{\mu\nu}$. Meanwhile, the $T_{ab}$ is the stress tensor of the matter field which lives on the spacetime boundary, i.e. $T_{ab}=-\frac{2}{\sqrt{-\gamma}}\frac{\delta S_{CFT}}{\delta \gamma^{ab}}$. Note that only the tangential components of the junction condition are nontrival, thus we multiply the vielbein $e^\mu _a$ in above equation. Actually, $\eqref{ExtenJuncCon}$ could be viewed as the generalised Israel junction condition for this type of vector-tensor theory $\eqref{EinProBHaction}$. The r.h.s. of $\eqref{ExtenJuncCon}$ is derived from the geometry of bulk spacetime, while the $T_{ab}$ in l.h.s corresponds to the stress tensor of the matter field living on the spacetime boundary. 

\subsection{Surface counterterms and holographic energy }

As indicated in \cite{Brown:1992br}, the total energy of black hole system, i.e. the black hole mass, is defined as a conserved charge associated with a timelike killing vector,
\begin{align}
\nonumber
&M=\int_{r\to\infty} d\Omega_2 \big\{ r^2\big(h(r)\big) ^{-\frac{1}{2}}T_{00} \big\}\\
\label{defiBHMass}
&\quad~=-\sqrt{5}\pi \frac{(Q^2_2-8)}{\Lambda} T_{00} \vert_{r\to\infty}
\end{align} 
where $T_{00}$ is the $00$-components of stress tensor \eqref{ExtenJuncCon}. However, the r.h.s of \eqref{ExtenJuncCon} typically diverges as $r\to\infty$. Specifically, substitute the solutions $\eqref{metrictt},\eqref{metricrr},\eqref{elect},\eqref{elecr}$ into \eqref{ExtenJuncCon} and then expand the results into power series, we could observe 
\begin{align}
T_{00}\vert_{r\to\infty}=c_3 r^3+c_1 r+c_0+O(\frac{1}{r})\dots
\end{align}
in which the coefficients $c_3, c_1$ are constituted by $Q_2$ and $\Lambda$, while the $c_0$ consist of $Q, Q_2, m, \Lambda$. 

From the viewpoint of gauge/gravity duality, the $T_{ab}$ is interpreted as the  expectation value of the stress tensor in the side of boundary field theory \cite{Witten:1998qj, Henningson:1998gx, Hyun:1998vg, Chalmers:1999gc, Nojiri:1998dh}. And the UV (ultraviolet) divergences appear naturally in field theory as the energy scale increases, which is reflected by the divergences of geometry quantity given by r.h.s of equation \eqref{ExtenJuncCon} as the boundary is taken to infinity. In renormalization procedure of quantum field theory, all UV divergences could be removed by adding the counterterms to the bare Lagrangian. In the same spirit for gravity theory, a finite set of boundary terms could be constructed as the counterterms which only cancel the divergences on the boundary spacetime and do not change the equations of motion in the bulk spacetime. 

But in our case, the surface counterterms will be more complicated than in \cite{Balasubramanian:1999re}, in which a standard expression of surface counterterms associated with a gravitational system in asymptotically AdS spacetime have been constructed. First, besides the gravitational field and negative cosmology constant, there also exists the vector field which couples to the gravitational field in nonminimal ways. Thus, except the intrinsic geometry quantities on the spacetime boundary, the surface counterterms should also include the contributions from the vector field. Secondly, the black hole solution \eqref{metrictt}-\eqref{metricrr} is not asymptotically AdS. So, the standard results of counterterms for $AdS_4$ given by \cite{Balasubramanian:1999re} are not applicable directly in the current case, and some non-standard results have to be developed.

Under $r\to\infty$ limit, for cancelling the divergences in stress tensor \eqref{ExtenJuncCon}, we construct the following counterterms
\begin{align}
\nonumber
&S_{ct}=\frac{-1}{\kappa^{2}}\int d^{3}x\sqrt{-\gamma}\big\{\frac{c_{0}}{\ell\sqrt{A^{\mu}A_{\mu}}}+\frac{c_{1}\ell\mathcal{R}}{\sqrt{A^{\mu}A_{\mu}}}+c_{m1}n^{\mu}F_{\mu\nu}A^{\nu}\\
\label{AdS4ProcaCounter}
&\quad \quad ~+c_{m2}\frac{n^{\mu}F_{\mu\nu}A^{\nu}}{A^{\alpha}A_{\alpha}}+c_{m3}\ell\frac{(n^{\mu}F_{\mu\nu}A^{\nu})^{2}}{\sqrt{A^{\alpha}A_{\alpha}}}\big\}
\end{align}
where $\mathcal{R}$ is the Ricci scalar of the induced metric $\gamma_{ab}$ and $\ell$ is AdS radius related with cosmological constant through $\Lambda=-\frac{3}{\ell^2}$. Note that the $c_0,c_1,c_{m1},c_{m2},c_{m3}$ are undetermined coefficients; they will be given later. With the inclusion of counterterms $\eqref{AdS4ProcaCounter}$, the stress tensor will be rewritten as
\begin{align}
\nonumber
&\hspace{-2mm}T^{eff}_{ab}=T^0_{ab}+\frac{c_{0}}{\ell}\big(\frac{\gamma_{ab}}{\sqrt{A^{\mu}A_{\mu}}}+\frac{e_{a}^{\alpha}e_{b}^{\beta}A_{\alpha}A_{\beta}}{\sqrt{(A^{\mu}A_{\mu})^{3}}}\big)+\ell c_{1}\big\{\frac{\gamma_{ab}\mathcal{R}}{\sqrt{A^{\mu}A_{\mu}}}\\
\nonumber
&\quad-\frac{2\mathcal{R}_{ab}}{\sqrt{A^{\mu}A_{\mu}}}+2(D_{a}D_{b}\frac{1}{\sqrt{A^{\mu}A_{\mu}}}-\gamma_{ab}D^{c}D_{c}\frac{1}{\sqrt{A^{\mu}A_{\mu}}})\\
\nonumber
&\quad+\frac{\mathcal{R}e_{a}^{\mu}e_{b}^{\nu}A_{\mu}A_{\nu}}{\sqrt{(A^{\mu}A_{\mu})^{3}}}\big\}+c_{m_1}\big\{\gamma_{ab}n^{\mu}F_{\mu\nu}A^{\nu}-2e_{(a}^{\nu}e_{b)}^{\beta}n^{\mu}F_{\mu\nu}A_{\beta}\big\}\\
\nonumber
&\quad+c_{m_2}\big\{\gamma_{ab}\frac{n^{\mu}F_{\mu\nu}A^{\nu}}{A^{\rho}A_{\rho}}+2e_{a}^{\beta}e_{b}^{\gamma}A_{\beta}A_{\gamma}\frac{n^{\mu}F_{\mu\nu}A^{\nu}}{(A^{\rho}A_{\rho})^{2}}\\
\nonumber
&\quad-2e_{(a}^{\nu}e_{b)}^{\beta}\frac{n^{\mu}F_{\mu\nu}A_{\beta}}{A^{\rho}A_{\rho}}\big\}+c_{m_{3}}\ell\big\{ e_{a}^{\rho}e_{b}^{\sigma}A_{\rho}A_{\sigma}\frac{(n^{\mu}F_{\mu\nu}A^{\nu})^{2}}{\sqrt{(A^{\alpha}A_{\alpha})^{3}}}\\
\label{stressEff}
&~~-n^{\mu}F_{\mu\nu}A_{\beta}e_{(a}^{\nu}e_{b)}^{\beta}\frac{4n^{\rho}F_{\rho\sigma}A^{\sigma}}{\sqrt{A^{\alpha}A_{\alpha}}}+\gamma_{ab}\frac{(n^{\mu}F_{\mu\nu}A^{\nu})^{2}}{\sqrt{A^{\alpha}A_{\alpha}}}\big\}
\end{align}
in which $D_a$ is the covariant differentiation with respect to induced metric $\gamma_{ab}$. Meanwhile, we use $T^0_{ab}$ to denote the bare stress tensor \eqref{ExtenJuncCon}. 

To remove the $O(r^3)$ and $O(r)$ divergences in the bare stress tensor $T^0_{ab}$, the undetermined coefficients are chosen as $c_{0}=\frac{32\sqrt{5}}{5\sqrt{3}}~,c_{1}=\frac{44}{9\sqrt{15}}~,c_{m_{1}}=-\frac{7}{8}~,c_{m_{2}}=-\frac{154}{15}~,c_{m_{3}}=-\frac{5\sqrt{5}}{32\sqrt{3}}$. Substitute the above coefficients and $\eqref{metrictt},\eqref{metricrr},\eqref{elect},\eqref{elecr},\eqref{ExtenJuncCon}$ into the $\eqref{stressEff}$, we obtain
\begin{align}
\label{RenorStress}
&T^{eff}_{00}\vert_{r\to\infty}=\frac{(5m(Q_{2}^{2}-8)-3QQ_{2})\Lambda}{2\sqrt{5}(Q_{2}^{2}-8)}+O(\frac{1}{r})+\dots
\end{align}
According to the above results and $\eqref{defiBHMass}$, the black hole mass is given as
\begin{align}
\label{EinProcaBHMass}
M=\frac{\pi}{2}\big(3QQ_{2}-5m(Q_{2}^{2}-8)\big)		
\end{align}

\section{Thermodynamics and phase structure\label{BHThermodynamic}}

\subsection{The first law of black hole thermodynamics}
From the metric ansatz $\eqref{MetAnsatz}$, the Hawking temperature could be caculated as
\begin{align}
\label{ProcaBHtem}
T=\frac{\sqrt{h^{\prime}f^{\prime}}}{4\pi}\big\vert_{r_{h}}=\frac{\vert4\Lambda r_{h}^{2}+Q_{2}^{2}-8\vert}{4\pi r_{h}(Q_{2}^{2}-8)}
\end{align}
As explained in Sec.\ref{AdSProBHSol}, we only consider $Q^2_2>8$ in this paper. For convenience in the following, we define $r_h ^\star$ as the zero-point of expression $\eqref{ProcaBHtem}$. And it is easy to see,
\begin{align}
\label{ProcaBHtemEx}
T=\begin{cases}
\begin{array}{c}
\frac{4\Lambda r_{h}^{2}+Q_{2}^{2}-8}{4\pi r_{h}(Q_{2}^{2}-8)}~,~r_{h}<r_{h}^{\star}\\
0~,~r_{h}=r_{h}^{\star}\\
-\frac{4\Lambda r_{h}^{2}+Q_{2}^{2}-8}{4\pi r_{h}(Q_{2}^{2}-8)}~,~r_{h}>r_{h}^{\star}
\end{array}\end{cases}		
\end{align}
According to the area law, the entropy of black hole is
\begin{align}
S=\frac{A}{4G}
\end{align}
in which $A$ represents the area of event horizon. Note that the convention $\kappa=8\pi G=1$ has been used in this paper, and the $S$ will be rewritten as
\begin{align}
\label{ProcaBHEntropy}
S=2\pi A=8\pi^{2} r_{h}^{2}
\end{align}
From the viewpoint of dynamics, the $Q$ is just an integration constant solved from the Einstein field equation and extended Proca equation. In $\eqref{elect}$, $\eqref{elecr}$, it seems that the value of $Q$ is free to choose and independent with other physical parameters like $Q_2, r_h, \Lambda$. However, for making the first law of black hole thermodynamics to hold,
\begin{align}
\label{BHFirstLaw}
dM=TdS+\Phi dQ
\end{align}
a constraint relation between $Q$ and $Q_2, r_h, \Lambda$ is implied,
\begin{align}
\nonumber
&Q=\frac{8r_{h}^{5}\Lambda^{2}}{3Q_{2}(Q_{2}^{2}-8)}+\frac{5(Q_{2}^{2}-8)r_{h}}{6Q_{2}}+\frac{20r_{h}^{3}\Lambda}{9Q_{2}}\\
\label{rhReplaQ}
&\quad+\begin{cases}
\begin{array}{c}
\frac{8\big(\frac{4}{3}\Lambda r_{h}^{3}+(Q_{2}^{2}-8)r_{h}\big)}{3Q_{2}(Q_{2}^{2}-8)}~,~r_{h}\leq r_{h}^{\star}\\
\frac{2C_{0}(r_{h}^{\star})}{3Q_{2}\pi}-\frac{8\big(\frac{4}{3}\Lambda r_{h}^{3}+(Q_{2}^{2}-8)r_{h}\big)}{3Q_{2}(Q_{2}^{2}-8)}~,~r_{h}\geq r_{h}^{\star}
\end{array}\end{cases}
\end{align}
in which the function $C_{0}(r_{h}^{\star})$ has the following expression,
\begin{align}
\nonumber
C_{0}(r_{h}^{\star})=\frac{8\pi\big(\frac{4}{3}\Lambda r_{h}^{\star3}+(Q_{2}^{2}-8)r_{h}^{\star}\big)}{(Q_{2}^{2}-8)}
\end{align} 

With thermodynamic quantities and constraint presented above, let us check that the $\eqref{BHFirstLaw}$ is established properly. Through the definition of horizon radius $h(r_h)=0$, a constraint relation between $m$ and $r_h, Q_2, \Lambda$ is deduced as follows
\begin{align}
\label{rhReplam}
m=\frac{r_{h}}{2}+\frac{4\Lambda r_{h}^{3}}{3(Q_{2}^{2}-8)}+\frac{8\Lambda^{2}r_{h}^{5}}{5(Q_{2}^{2}-8)^{2}}
\end{align}
Substitute $\eqref{rhReplaQ}$ and $\eqref{rhReplam}$ into $\eqref{EinProcaBHMass}$, we get
\begin{align}
\label{EinProBHMRepm}
&\hspace{-2.2mm}M=\begin{cases}
\begin{array}{c}
\frac{4\pi\big(\frac{4}{3}\Lambda r_{h}^{3}+(Q_{2}^{2}-8)r_{h}\big)}{(Q_{2}^{2}-8)}~,~r_{h}\leq r_{h}^{\star}\\
-\frac{4\pi\big(\frac{4}{3}\Lambda r_{h}^{3}+(Q_{2}^{2}-8)r_{h}\big)}{(Q_{2}^{2}-8)}+C_{0}(r_{h}^{\star})~,~r_{h}\ge r_{h}^{\star}
\end{array}\end{cases}		
\end{align}
Combine $\eqref{EinProBHMRepm}$ with $\eqref{ProcaBHtemEx}$, $\eqref{ProcaBHEntropy}$, it is easy to check 
\begin{align}
\frac{\partial M}{\partial S}=\big(\frac{\partial M}{\partial r_h}\big) \bigg/\big(\frac{\partial S}{\partial r_h}\big)=T
\end{align}
Due to the breaking of $U(1)$ symmetry in Einstein-Proca theory, the electric charge is not the locally-conserved charge any more. And we could not obtain the electric charge directly from the Gauss law which doesn't hold in current situation. However, from the solution of electric potential $\eqref{elect}$, it is not diffcult to observe that the $Q$ and $Q_2$ play the roles of charge and chemical potential respectively. In other way, from black hole mass $\eqref{EinProcaBHMass}$, we could obtain the following relation directly,
\begin{align}
\frac{\partial M}{\partial Q}=\frac{3\pi}{2} Q_2
\end{align}
it also means that the $Q$ and $Q_2$ are a pair of conjugate variables in thermodynamics. Exactly, we define the chemical potential $\Phi$ as 
\begin{align}
\label{ChemiPotential}
\Phi=\frac{3\pi}{2} Q_2
\end{align}
Note that the charge is not the conserved quantity in current situation, thus we consider the thermodynamics in the grand canonical ensemble.

\begin{figure}[ht]
	\begin{center}
		\includegraphics[scale=0.4]{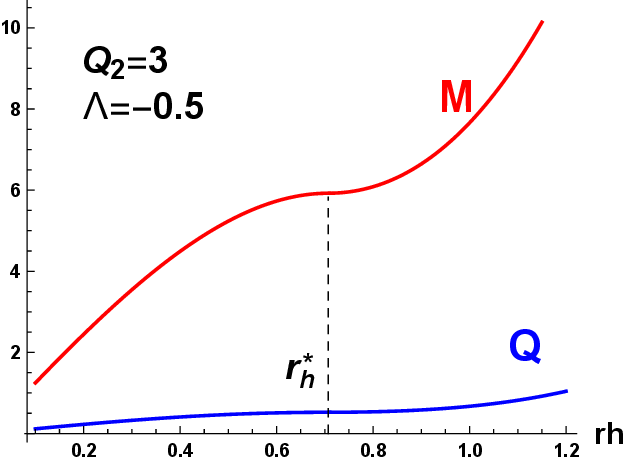}
		\includegraphics[scale=0.4]{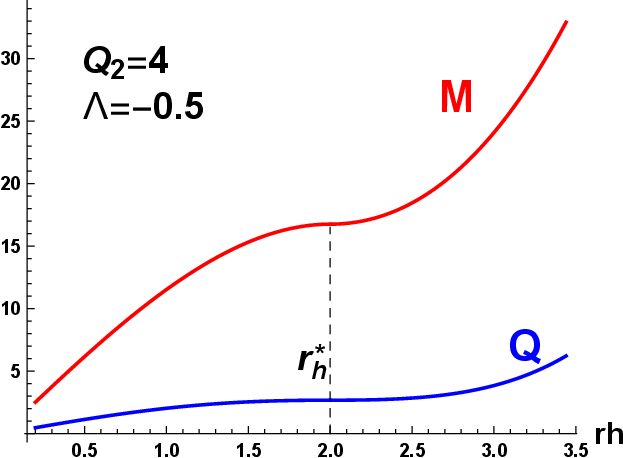}\\
		\includegraphics[scale=0.4]{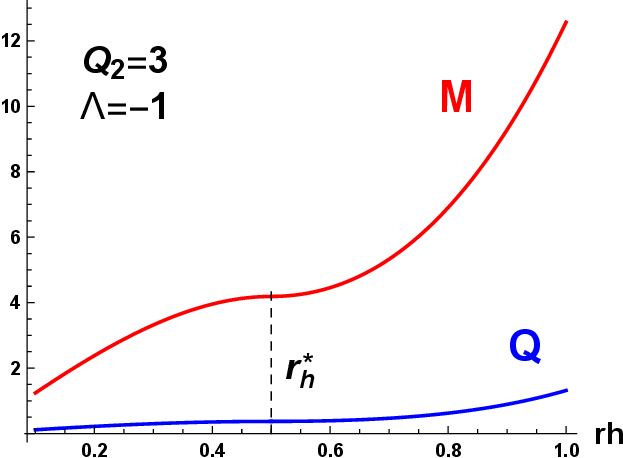}
		\includegraphics[scale=0.4]{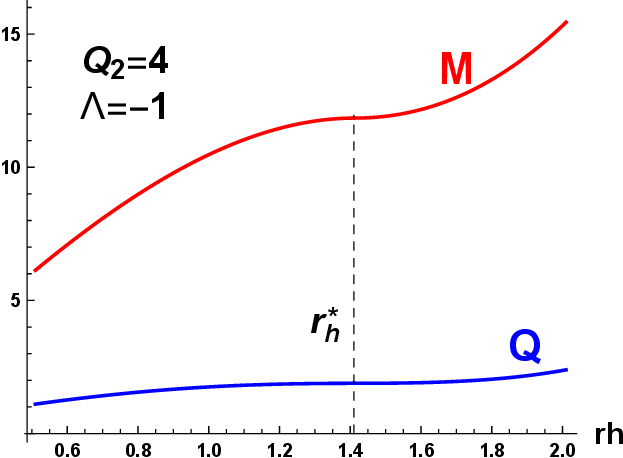}
		\caption{(color online).Compare charge and black hole mass in some representative parameters.}
		\label{MvsQrh}
	\end{center}
\end{figure}

As we known, for the  Reissner-Nordstr$\ddot{o}$m black hole solved from Einstein-Maxwell theory,  $Q\leq M$, or the horizon does not exist. More generally, this extremality bound holds for most of charged black holes. From the black hole solution $\eqref{metrictt}$-$\eqref{elecr}$, it seems that the value of $Q$ is arbitrary and has no effect on the existence of horizon. However, after the consideration of thermodynamics, we find a constraint relation between $Q$ and other physical parameters, on which the first law of black hole thermodynamics could be reproduced correctly. Finally, by using $\eqref{rhReplaQ}$ and $\eqref{EinProBHMRepm}$ implied by thermodynamics, we can compare charge $Q$ and black hole mass $M$ at fixed horizon $r_h$, chemical potential $\Phi (Q_2)$ and cosmological constant $\Lambda$. As shown by Fig.\ref{MvsQrh}, we observe that the value of $Q$ less than the value of $M$ indeed.

\subsection{Phase structures analysis}

\begin{figure}[ht]
	\begin{center}
		\includegraphics[scale=0.68]{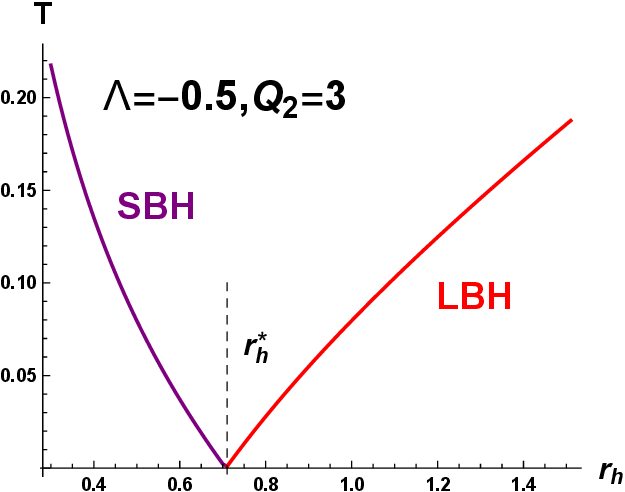}
		\caption{(color online).The variation of temperature versus the horizon radius $r_h$ for the four-dimensional black hole solutions with spherical horizon, in representative parameters $\Lambda=-0.5$ and $Q_2 =3$.}
		\label{BHPhase}
	\end{center}
\end{figure}

According to the expression of the Hawking temperature $\eqref{ProcaBHtem}$, we plot the variation of temperature versus the horizon radius $r_h$ in Figure \ref{BHPhase}. For a given temperature $T>0$, there exists two values of horizon $r_h$. We call them "small black hole" (SBH) and "large black hole" (LBH) phases, as shown by purple and red curves in Figure \ref{BHPhase} respectively.

\begin{figure}[ht]
	\begin{center}
		\includegraphics[scale=0.67]{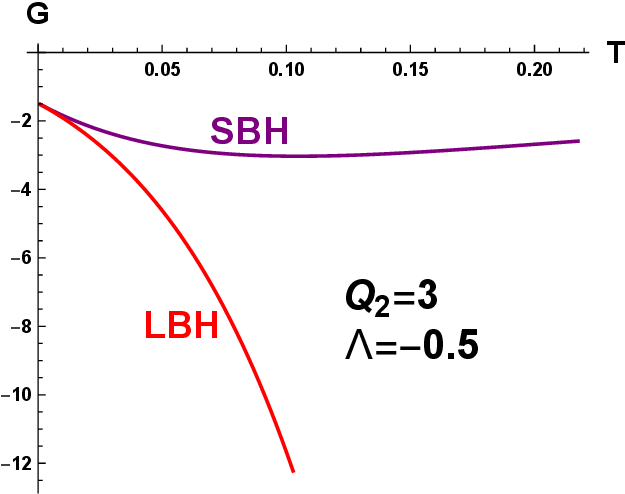}
		\caption{(color online).The variation of Gibbs free energy $G$ versus the Hawking temperature $T$ for the four-dimensional black hole solutions with spherical horizon, in representative parameters $\Lambda=-0.5$ and $Q_2 =3$.}
		\label{BHGTdiagram}
	\end{center}
\end{figure}

Note that the SBH and LBH coexist at the same temperature, and we need to determine which phase is thermodynamically preferred. This goal could be achieved by comparing the Gibbs free energy of SBH and LBH; the one which has lower free enegy will be favored by the thermodynamical system. In the grand canonical ensemble, the Gibbs free energy is defined as
\begin{align}
\label{DefineGibbsGran}
&G=M-TS-\Phi Q
\end{align}
substitute $\eqref{ProcaBHtemEx}$, $\eqref{ProcaBHEntropy}$, $\eqref{rhReplaQ}$, $\eqref{EinProBHMRepm}$, $\eqref{ChemiPotential}$ into $\eqref{DefineGibbsGran}$, we have
\begin{align}
\nonumber
&G=-\frac{5}{4}(Q_{2}^{2}-8)\pi r_{h}-\frac{10}{3}\pi r_{h}^{3}\Lambda-\frac{4\pi r_{h}^{5}\Lambda^{2}}{Q_{2}^{2}-8}\\
\label{SpeficGibbs}
&\quad\quad-2\pi r_{h}\times\begin{cases}
\begin{array}{c}
\frac{4\Lambda r_{h}^{2}+Q_{2}^{2}-8}{(Q_{2}^{2}-8)}~,~r_{h}\leq r_{h}^{\star}\\
-\frac{4\Lambda r_{h}^{2}+Q_{2}^{2}-8}{(Q_{2}^{2}-8)}~,~r_{h}\ge r_{h}^{\star}
\end{array}\end{cases}
\end{align}
In Figure \ref{BHGTdiagram}, we show that how Gibbs free energy $G$ varies with respect to the temperature $T$. It is easy to observe that the Gibbs free energy of LBH is lower than the one of SBH, thus the LBH is thermodynamically prefered.
A similiar conclusion could also be supported on comparing the specific heat of SBH and LBH. In the  grand canonical ensemble, the specific heat is defined as
\begin{align}
\label{SpeHeatGrand}
&C_{\Phi}=T\big(\frac{\partial S}{\partial T}\big)_{\Phi}=T\bigg(\frac{\partial S/\partial r_{h}}{\partial T/\partial r_{h}}\bigg)_{\Phi}
\end{align}
substitute $\eqref{ProcaBHtem}$, $\eqref{ProcaBHEntropy}$ into $\eqref{SpeHeatGrand}$ and expand it explicitly, we obtain
\begin{align}
&C_{\Phi}=\frac{16\pi^{2}r_{h}^{2}(8-Q_{2}^{2}-4r_{h}^{2}\Lambda)}{Q_{2}^{2}-8-4r_{h}^{2}\Lambda}
\end{align}
\begin{figure}[ht]
	\begin{center}
		\includegraphics[scale=0.67]{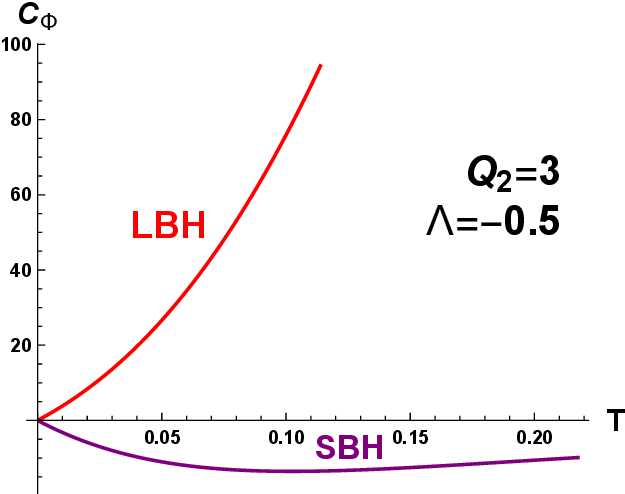}
		\caption{(color online).The variation of specific heat $C_\Phi$ versus the Hawking temperature $T$ for the four-dimensional black hole solutions with spherical horizon, in representative parameters $\Lambda=-0.5$ and $Q_2 =3$.}
		\label{BHCTdiagram}
	\end{center}
\end{figure}
From the $C_\Phi-T$ diagram, we easily see that the LBH phase has the positive specific heat, so it is thermodynamically stable. By contrast, the SBH is unstable.

Actually, the unstable SBH phase and stable LBH phase also imply a self-consistent result for thermodynamic quantities in extended phase space. In $\eqref{EinProBHMRepm}$, we could identify that the branch in region $r_h\leq r_h ^\star$ and $r_h\geq r_h ^\star$ correspond to the SBH phase and LBH phase respectively. If we consider the black hole thermodynamics in extend phase space~\cite{Kastor:2009wy, Kubiznak:2012wp}, and define the black hole pressure $P$ as,
\begin{align}
&P=-2\pi\Lambda_{eff}=-\frac{4\pi \Lambda}{Q^2_2-8}
\end{align}
Then, from $\eqref{EinProBHMRepm}$, we can get
\begin{align}
\label{VoluInExtend}
\partial M\big/\partial P=\begin{cases}
\begin{array}{c}
-\frac{4}{3}\pi r_{h}^{3}~,~r_h\leq r^\star_h\\
\frac{4}{3}\pi r_{h}^{3}~,~r_h\geq r^\star_h
\end{array}\end{cases}
\end{align}
In LBH phase, the conjugate variables for pressure $P$ is exactly the black hole volume. But the result is unphysical in SBH phase. Thus we naturally expect that the SBH phase is not allowed by the thermodynamic system, in other words, this phase is unstable.

\section{Conclusions and Discussion \label{ConDiss}}

For the type of non-minimally coupled vector-tensor theory given in \cite{Chagoya:2016aar}, we computed and added the appropriate boundary terms $\eqref{TotalBou}$ to the bulk action $\eqref{EinProBHaction}$ and derived the correspondingly well-defined junction condition $\eqref{ExtenJuncCon}$ on the spacetime boundary. As interpreted by the gauge/gravity duality, the $T_{ab}$ in $\eqref{ExtenJuncCon}$ represents the stress tensor of matter fields on the spacetime boundary. Meanwhile, this stress tensor is a divergent quantity. For removing the divergences in the stress tensor,  surface counterterms need to be added to the total action. A standard expression of surface counterterms associated with a gravitational system in asymptotically AdS spacetime has been constructed by \cite{Balasubramanian:1999re, Emparan:1999pm}, in which the surface counterterms consist of the cosmological constant and intrinsic geometry quantities of the bounday spacetime like the Ricci scalar and Ricci tensor for the boundary metric $\gamma_{ab}$. However, the surface counterterms in our case are more complicate than the results in \cite{Balasubramanian:1999re, Emparan:1999pm}, even though in four-dimensional spacetime. Our major difficulties come from the following two aspects. On the one hand, the gauge symmetry is broken due to the existence of nonminimal coupling between the vector field and Einstein tensor. On the other hand, the black hole solutions $\eqref{metrictt}$-$\eqref{metricrr}$ are not an asymptotic AdS geometry as shown by $\eqref{AsyLarger}$. Thus, unlike the expression of surface counterterms given by \cite{Balasubramanian:1999re, Emparan:1999pm}, the mixture of vector field and intrinsic geometry quantities of bounday spacetime need to be considered in our case. Finally, after various attempts, we find the surface counterterms $\eqref{AdS4ProcaCounter}$. Based on $\eqref{AdS4ProcaCounter}$, we obtain the finite stress tensor $\eqref{RenorStress}$ on boundary spacetime and calculate the corresponding conserved charge associated with a timelike killing vector, i.e. the black hole mass $\eqref{EinProcaBHMass}$. 

From $\eqref{metrictt}$-$\eqref{elecr}$, it is easy to see that the value of $Q$ is independent of other parameters and has no effect on the existence of a horizon. Thus, both $Q>M$ and $Q\leq M$ are allowed according to the expression $\eqref{EinProcaBHMass}$. However, for most of charged black hole, there exists a general extremality bound $\frac{Q}{M}\leq 1$. Actually, after consideration of the first law of black hole thermodynamics, an implicit relation of $Q$ as the function of $Q_2$, $\Lambda$ and horizon $r_h$ are found in $\eqref{rhReplaQ}$. Then, by using $\eqref{rhReplaQ}$ and $\eqref{EinProBHMRepm}$, we can compare the charge $Q$ with black hole mass $M$ directly at fixed $Q_2$, $\Lambda$ and $r_h$. As shown by Fig.\ref{MvsQrh}, we indeed observe that the value of $Q$ less than the value of $M$. Besides, after the analysis of phase structure in grand canonical ensemble, we observe the unstabel SBH phase and stable LBH as displayed by Fig.\ref{BHPhase}-Fig.\ref{BHCTdiagram}. In fact, as we show in $\eqref{VoluInExtend}$, the unstable SBH phase and stable LBH phase also imply a self-consistent result for well-defined pressure and volume in extended phase space.

As future research, we suggest the following extended topics. In  \cite{Golovnev:2008cf}, inflation is driven by non-minimally coupled massive vector fields. It would be interesting consider cosmic inflation by using the vector-tensor theory $\eqref{EinProBHaction}$. Besides, it is also worthwhile to develop the supersymmetric extension of $\eqref{EinProBHaction}$ and investigate the relevant phenomenology in black hole physics and cosmology.

\section{Acknowledgements}
We would like to thank C. Herdeiro and E. Radu, Ru-yong Li for fruitful and useful discussions. This work is supported by The Center for Research and Development in Mathematics and Applications (CIDMA) through the Portuguese 
Foundation for Science and Technology (FCT - Fundação para a Ciência e a Tecnologia), references UIDB/04106/2020 and UIDP/04106/2020. Besides, A.-c.L is also partly supported by NSFC grant no.11875082.

\end{document}